\documentclass[journal,onecolumn,10pt]{IEEEtran}
\usepackage{setspace}
\begin{document}

\title{Deep Layered LMS Predictor}

\author{Lubna~Shibly~Mokatren,~\IEEEmembership{Student Member,~IEEE,}
        Ahmet~Enis~Cetin,~\IEEEmembership{Member,~IEEE,}
        and Rashid~Ansari,~\IEEEmembership{Member,~IEEE,}
        \thanks{L. Shibly Mokatren, R. Ansari and A. Cetin are with the Department
of Electrical and Computer Engineering, University of Illinois at Chicago, Chicago}}

\maketitle

\begin{abstract}
In this study, we present a new approach to design a Least Mean Squares (LMS) predictor. This approach exploits the concept of deep neural networks and their supremacy in terms of performance and accuracy. The new LMS predictor is implemented as a deep neural network using multiple non linear LMS filters. The network consists of multiple layers with nonlinear activation functions, where each neuron in the hidden layers corresponds to a certain FIR filter output which goes through nonlinearity. The output of the last layer is the prediction. We hypothesize that this approach will outperform the traditional adaptive filters.

\end{abstract}

\begin{IEEEkeywords}
LMS, adaptive filters, deep learning
\end{IEEEkeywords}

\IEEEpeerreviewmaketitle
\section{Introduction}

\IEEEPARstart{A}{daptive} filtering is a very significant topic that is used in a wide range of applications, such as bio-medical systems and digital communications. Adaptive filters play an important role in signal processing products (DSP) due to their great practical and theoretical value \cite{Sayed:2008:AF:1370975}. The filters deal with problems in many areas with changing environment such as noise cancellation, channel equalization, system identification, prediction, and inverse modeling. 
An adaptive filter is a closed loop filter that has self-adjusting characteristic according to an optimization algorithm to obtain the best desired signal in spite of the changing conditions. Adaptive filters work generally for the adaptation of a certain signal and its efficiency depends on the design technique used and the algorithm of adaptation \cite{sayed2011adaptive},\cite{bellanger2001adaptive}. The characteristics of adaptive algorithms make them very popular in signal processing field and significant in a variety of applications, such as image compression \cite{gerek2000adaptive}, \cite{gerek1998polyphase}, detection in medical images \cite{gurcan1998microcalcifications}, robustness of filtering in impulsive noise environments \cite{aydin1999robust},  and sound reproduction systems \cite{bouchard2000multichannel}. Most of the adaptive filters are digital due to the complexity of the optimization algorithms. 

We can look at the adaptive filter as a device in which a digital input signal x(n) is fed into it, and it computes the output y(n) at time instance n. We define the error signal e(n) at time n, as the difference between the output signal y(n) and the desired signal d(n). This error signal is fed to the procedure that adjust the filter parameters at time n+1. After some time of performing the adaptation process, the parameters are adapted to better match the desired output signal. The adaptive filter algorithms discussed in this paper are implemented with finite impulse response (FIR) filter structures. The two most common adaptive algorithms are known as Least Mean Squares (LMS) and Recursive Least Squares (RLS). In this study, our main focus is LMS adaptive algorithms.

The Least Mean Square (LMS) adaptive filter was first devised by Widrow and Hoff in 1959 during their study of a pattern recognition machine known as Adaline \cite{haykin2003least}. LMS is the core algorithm that is most referenced in adaptive filters, and is successfully used in a wide range of applications \cite{Sayed:2008:AF:1370975}, 
such as image compression \cite{oktem2001lossless}, video coding \cite{toreyin2006lms}, identification of a linear system \cite{alshebeili1991adaptive}, and wildfire detection \cite{toreyin2009wildfire}. It is very popular in many implementations due to its robustness to signal information, simplicity and low computational complexity \cite{sayed1996error}. However, its convergence speed might be slow depending on the statistics of the input signal. 

In general, the closed loop adaptive process of the adaptive filter involves the use of a cost function as a feedback for optimum performance.  During this process, the filter is adjusted until the error e(n) is minimized. e(n) is defined as the difference between the output signal y(n) and the desired signal d(n) at time n. 
\\The structure of the weights or parameters of FIR filter can be estimated using the coefficients vector \cite{kolinova1998adaptive}.
The weight vector of the filter is defined as follows:
 \begin {equation}
\bar{w}(n)=[ w_0(n),w_1(n),...,w_L(n)]
\end {equation}
 where L is the order of the filter and n is the time instance. The input vector can be defined as :
  \begin {equation}
 \bar{x}(n)=[ x(n),x(n-1),...,x(n-L)]
  \end {equation}
  The output of the filter is y(n) :
    \begin {equation}
 y(n)=\bar{w}^T(n)\bar{x}(n)
  \end {equation}
 
The weight update equation for the filter is derived through a minimization of the cost function defined as :
\begin {equation}
 min E[e(n)^2]= min E[(d(n)-y(n))^2]
  \end {equation}
The Wiener filter coefficients are obtained by minimizing the cost function with respect to the filter weight vector w(n). The weight update is performed using gradient descent technique, which uses the gradients of the performance surface in seeking its minimum. 
 
In this study we will consider the prediction version of the LMS adaptive algorithm.  A computationally simpler version of the gradient search method is the LMS in which the gradient of the mean square error is substituted with the gradient of the squared error or stochastic gradient. The LMS adaptation equation is defined as follows:
\begin {equation}
\bar{w}(n+1)=\bar{w}(n)+{\mu}e(n)\bar{x}(n)  
  \end {equation}
After the weights at time n are adjusted, the next set of weights, i.e. $\bar{w}(n+1)=[ w_0(n+1),w_1(n+1),...,w_L(n+1)]$ are used to estimate the output y(n+1) in an iterative manner.

A main problem in the LMS algorithm is that it is sensitive to the scaling of its input \cite{dhiman2013comparison}. Hence, it makes the choosing of a suitable learning rate that guarantees stability more difficult. The Normalized LMS (NLMS) algorithm is a modified form of the standard LMS algorithm that solves this problem \cite{haykin1999adaptive}, \cite{chen2003proportionate}. It is obtained from Eq. (5) by normalizing the update portion of the equation using the Euclidian norm of vector x(n):
\begin {equation}
\bar{w}(n+1)=\bar{w}(n)+{\lambda}\frac{e(n)}{\left \| x \right \|^{2}}\bar{x}(n)  
  \end {equation}

The upper bound on $\lambda$ to ensure convergence is 2. The NLMS algorithm converges to the Wiener filter solution of the linear minimum mean square error estimator LMMSE predictor design when $0<\lambda<2$  and x[n] is a wide sense stationary random process. If the random process is not WSS the NLMS tracks the error by performing an orthogonal projection onto the hyperplane to determine the next set of filter weights. 
The NLMS is easy linear filter to implement, however it is slow to converge. Another popular adaptive predictor is based on the Recursive least squares (RLS) algorithm. RLS converges much faster than the LMS adaptive filter algorithm but its tracking performance is not as good as the LMS algorithm when the data is non-stationary. One way to implement NLMS is using a simple single layer neural network, however it does not really enhance the performance by much.  Different implementations of LMS algorithm and similar linear algorithms are still very limited in many applications based on their performances.
Deep learning has significantly improved the performance for many problems compared with other machine learning algorithms \cite{deng2014deep}. Hence, developing deep network-based filters can be very effective in many applications. 
We introduce in this study a new novel approach to design the NLMS predictor, which has a layered structure similar to the deep neural networks. This structure exploits the nonlinearities of the activation functions in order to enhance the prediction. 
 
 \section{method}
 
Based on the background above, the FIR filter has a single layer in the LMS algorithm as $ y(n)=\bar{w}^T(n)\bar{x}(n)$ where the input is $ \bar{x}(n)=[ x(n),x(n-1),...,x(n-L)]$ and w(n) are the weights of the different edges of the network, so we have one layer with L neurons.  A  novel predictor which has a layered structure is suggested. Instead of building the LMS as a single layer network with linear activation function, more layers and filters can be added to obtain a deep layered predictor. 
To demonstrate, we use a time series data called d, we can use the filter to predict the value of the signal at time n (d(n)) using the previous signals as follows:

  \begin {equation}
 \hat {d}(n)=[ w_1(n)d(n-1)+ w_2(n)d(n-2)+...+w_L(n)d(n-L)]  
   \end {equation}	

Where $\hat d(n)$ is the predicted or estimated signal at time n.  Assuming we have the data signals at time [n-1,n-2,...n-L], we add a bias DC term $b_0$ to get the input $\bar{x}(n)=[ d(n-1),...,d(n-L),b_0]$. Applying the LMS weights which are  $\bar{w}(n)=[ w_1(n),...,w_L(n),w_0(n)]$, we get the following output:

\begin {equation}
 \hat {d}(n)=\sum_{i=1}^{L} w_i(n)d(n-i)+w_0(n)b_0   
\end {equation}

Rather than traditional LMS filter implementation, this some will go through a nonlinearity such as a Rectified Linear Unit (RELU), where RELU(x) = max(x,0). Hence we obtain the following :
\begin {equation}
 a(n)=RELU(\sum_{i=1}^{L} w_i(n)d(n-i)+w_0(n)b_0)
\end {equation}
Here, a(n) indicates the prediction output of a nonlinear unit. In addition, we can repeat the previous steps similary and use more FIR filters. Let $\bar{v}(n)=[ v_1(n),...,v_L(n),v_0(n)]$
be the weights of the second filter, which are used to estimate a(n-1), which is the output of the second nonlinear unit based on the previous data:
\begin {equation}
 a(n-1)=RELU(\sum_{i=1}^{L} v_i(n)d(n-i)+v_0(n)b_0)
\end {equation}
More filters are used to generate another time series data [a(n),a(n-1),...] from the input $\bar{x}(n)=[d(n-1),...,d(n-L),b_0]$. We can have as many layers as we can similar to the deep neural networks structure with different activation functions or without applying any nonlinearities on some of the layers. At the last layer, we will have the final predictor:

\begin {equation}
 \hat {d}(n)=\sum_{i}h_i(n)a(n-i)+h_0(n)s_0   
\end {equation}
Here, the time series a, represents the values of the output neurons at the layer before last, and not necessary the second layer. It should be noted that various numbers of layers, type of nonlinearities and filter order can be used in the deep neural network.
The error at time n is defined as : $e(n)=d(n)- \hat {d}(n)$.
Once e[n] is available, the weights of the filters are updated using the gradient descent approach which is the standard back-propagation algorithm.

\section{results}
We hypothesize that this approach will outperform the traditional adaptive filters. Experiments are still being conducted.
One possible advantage of this approach is that it may automatically solve the order problem of FIR filtering methods \cite{kumar2015optimal}.  Orthogonal projection operation in the NLMS uniformly distributes the error onto weights.  As a result the performance of the LMS filter depends on the order of the filter L.  Multilayered structure may avoid the uniform distribution. Another advantage may be the robustness against noise or sudden spikes in data. Due to nonlinearities, a sudden unusual increase may be handled by the multilayered structure. 
We will also implement the RLS filter and compare its performance to the LMS algorithm.

\bibliography{IEEEabrv,shibly}
\bibliographystyle{IEEEtran}

\end{document}